\documentclass[twoside,twocolumn,9pt]{article}

\usepackage[utf8]{inputenc}
\usepackage{t1enc,latexsym,amsfonts,amssymb,amsmath}
\usepackage{bm}
\usepackage{graphics}
\usepackage{setspace}
\usepackage{url}

\usepackage{extsizes}
\usepackage[super,sort&compress,comma]{natbib}
\usepackage[version=3]{mhchem}
\usepackage[left=1.5cm, right=1.5cm, top=1.785cm, bottom=2.0cm]{geometry}
\usepackage{balance}
\usepackage{times,mathptmx}
\usepackage{sectsty}
\usepackage{graphicx}

\usepackage{lastpage}
\usepackage[format=plain,justification=justified,singlelinecheck=false,font={stretch=1.125,small,sf},labelfont=bf,labelsep=space]{caption}
\usepackage{float}
\usepackage{fancyhdr}
\usepackage{fnpos}
\usepackage[english]{babel}
\usepackage{array}
\usepackage{droidsans}
\usepackage{charter}
\usepackage[T1]{fontenc}
\usepackage[usenames,dvipsnames]{xcolor}
\usepackage{setspace}
\usepackage[compact]{titlesec}
\usepackage{color}

\usepackage{epstopdf}

\usepackage{dcolumn}   

\definecolor{cream}{RGB}{222,217,201}

\begin{document}

\pagestyle{fancy}
\thispagestyle{plain}
\fancypagestyle{plain}{

\fancyhead[C]{\includegraphics[width=18.5cm]{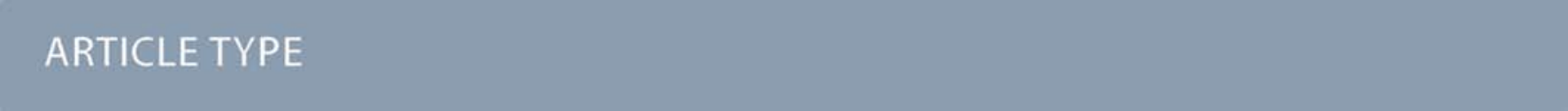}}
\renewcommand{\headrulewidth}{0pt}
}

\makeFNbottom
\makeatletter
\renewcommand\LARGE{\@setfontsize\LARGE{15pt}{17}}
\renewcommand\Large{\@setfontsize\Large{12pt}{14}}
\renewcommand\large{\@setfontsize\large{10pt}{12}}
\renewcommand\footnotesize{\@setfontsize\footnotesize{7pt}{10}}
\makeatother

\renewcommand{\thefootnote}{\fnsymbol{footnote}}
\renewcommand\footnoterule{\vspace*{1pt}%
\color{cream}\hrule width 3.5in height 0.4pt \color{black}\vspace*{5pt}}
\setcounter{secnumdepth}{5}

\makeatletter
\renewcommand\@biblabel[1]{#1}
\renewcommand\@makefntext[1]%
{\noindent\makebox[0pt][r]{\@thefnmark\,}#1}
\makeatother
\renewcommand{\figurename}{\small{Fig.}~}
\sectionfont{\sffamily\Large}
\subsectionfont{\normalsize}
\subsubsectionfont{\bf}
\setstretch{1.125} 
\setlength{\skip\footins}{0.8cm}
\setlength{\footnotesep}{0.25cm}
\setlength{\jot}{10pt}
\titlespacing*{\section}{0pt}{4pt}{4pt}
\titlespacing*{\subsection}{0pt}{15pt}{1pt}

\fancyfoot{}
\fancyhead{}
\renewcommand{\headrulewidth}{0pt}
\renewcommand{\footrulewidth}{0pt}
\setlength{\arrayrulewidth}{1pt}
\setlength{\columnsep}{6.5mm}
\setlength\bibsep{1pt}

\makeatletter
\newlength{\figrulesep}
\setlength{\figrulesep}{0.5\textfloatsep}

\newcommand{\topfigrule}{\vspace*{-1pt}%
\noindent{\color{cream}\rule[-\figrulesep]{\columnwidth}{1.5pt}} }

\newcommand{\botfigrule}{\vspace*{-2pt}%
\noindent{\color{cream}\rule[\figrulesep]{\columnwidth}{1.5pt}} }

\newcommand{\dblfigrule}{\vspace*{-1pt}%
\noindent{\color{cream}\rule[-\figrulesep]{\textwidth}{1.5pt}} }

\makeatother


\newcommand{\argc}[1]{\left[#1\right]}
\newcommand{\arga}[1]{\left\lbrace #1\right\rbrace }
\newcommand{\argp}[1]{\left(#1\right)}
\newcommand{\valabs}[1]{\vert #1\vert}
\newcommand{\moy}[1]{\left\langle  #1 \right\rangle }
\newcommand{\moydes}[1]{\overline{#1}}
\renewcommand{\vec}[1]{{\bf #1}}
\newcommand{\meevid}[1]{\textcolor{VioletRed}{#1}}
\newcommand{\comment}[1]{\textcolor{OliveGreen}{#1}}
\newcommand{\KM}[1]{\textcolor{blue}{Comment KM: #1}}
\newcommand{\LC}[1]{\textcolor{red}{Comment LC: #1}}
\newcommand{\LR}[1]{\textcolor{Green}{Comment LR: #1}}



\twocolumn[
  \begin{@twocolumnfalse}
\vspace{3cm}
\sffamily
\begin{tabular}{m{4.5cm} p{13.5cm} }

& \noindent\LARGE{\textbf{Microscopic precursors of failure in soft matter}} \\
\vspace{0.3cm} & \vspace{0.3cm} \\

 & \noindent\large{Luca Cipelletti$^{\ast}$\textit{$^{a}$}, Kirsten Martens\textit{$^{b}$} and Laurence Ramos\textit{$^{a}$}} \\
 & \noindent\normalsize{The mechanical properties of soft matter are of great importance in countless applications, in addition of being an active field of academic research. Given the relative ease with which soft materials can be deformed, their non-linear behavior is of particular relevance. Large loads eventually result in material failure. In this Perspective article, we discuss recent work aiming at detecting precursors of failure by scrutinizing the microscopic structure and dynamics of soft systems under various conditions of loading. In particular, we show that the microscopic dynamics is a powerful indicator of the ultimate fate of soft materials, capable of unveiling precursors of failure up to thousands of seconds before any macroscopic sign of weakening.} \\

\end{tabular}

 \end{@twocolumnfalse} \vspace{0.6cm}

  ]



\renewcommand*\rmdefault{bch}\normalfont\upshape
\rmfamily
\section*{}
\vspace{-1cm}


\footnotetext{\textit{$^{*}$~E-mail:} luca.cipelletti@umontpellier.fr}
\footnotetext{\textit{$^{a}$~L2C, Univ. Montpellier, 34095 Montpellier, France}}
\footnotetext{\textit{$^{b}$~Univ.~Grenoble Alpes, CNRS, LIPhy, 38000 Grenoble, France.}}


\section{Introduction}
\label{sec:intro}

The term `soft matter' was introduced more than forty years ago to emphasize how easily soft materials can be deformed. [Incidentally, the incipt of the \textit{Celebrating Soft Matter’s $15^{th}$ anniversary} Editorial~\cite{_celebrating_2019} should be amended: the expression was coined by Madeleine Veyssi\'{e}~\cite{book_demain_physique}, at that time professor at Paris Orsay and a collaborator of de Gennes, not by de Gennes himself. Ironically, the expression was born as a private joke, with ---in its French version, \textit{mati\`{e}re molle}--- a slightly  scatological twist, and initially de Gennes was not fond of it~\cite{note_mad}]. 
Because of the high responsiveness of soft materials to mechanical loading, the rheological behavior of soft matter is  at the very heart of this research field. The articles published in the last 15 years in \textit{Soft Matter} reflect its importance: a search for ``rheology'' or ``mechanics'' returns about 6800 items~\cite{note}, more than half of the articles published to date in this journal.

Upon mechanical loading, soft matter initially deforms according to the laws of linear (visco)elasticity. For larger loads, the system response departs from linearity, and at large enough load the material eventually fails. Failure in soft matter may occur through a variety of mechanisms, such as defect proliferation and migration in crystals, catastrophic strand rupture in network-forming systems, fracture or band nucleation in dense systems. In this Perspective article, we shall be concerned with amorphous viscoelastic materials and define quite broadly failure as the (often abrupt) transition between a predominantly solid-like response to flow. In experiments where a load is imposed, failure manifests itself as the significant increase of the resulting deformation or deformation rate. Conversely, upon imposing a controlled deformation or deformation rate, failure is signalled by a significant drop of the resistance force exerted by the sample.

Failure may occur almost instantaneously as the result of an impulsive load, think of a piece of glass shattered into pieces by a sudden impact. Perhaps more intriguingly, failure may also occur with no significant changes of the loading conditions, after an induction time that may last up to tens of hours, seemingly limited only by the researcher's patience. This \textit{delayed} failure is often elusive, since little (if any) precursor signs pinpointing the eventual fate of the material can be detected by monitoring conventional macroscopic quantities, such as those recorded in a rheological test.  Our viewpoint is that delayed failure is preceded by \textit{microscopic} precursors, as demonstrated by recent work that we shall briefly discuss in the following: we believe that this is a new exciting research area that will lead to a better understanding of the origin of failure, as well as provide new predictive tools.

What exactly do we mean by a microscopic precursor of failure? Once again, we shall take a broad definition and consider, for soft matter, any sharp variation of a parameter quantifying the system structure or dynamics on length scales of the order of, or slightly larger than, the relevant length scale of the sample structure, e.g. the particle size for colloids or the mesh size for network-forming systems. This signature should be detectable well in advance of macroscopic failure. Although mechanical tests often probe soft systems under shear, we do not restrict ourselves to a particular deformation geometry. Moreover,  we shall consider different kinds of rheology experiments, such as creep (constant imposed stress), stress relaxation (constant imposed strain), deformation startup (constant deformation rate), fatigue (many cycles of an imposed sinusoidal stress or strain).

\begin{figure}
    \centering
    \includegraphics[width=1\columnwidth]{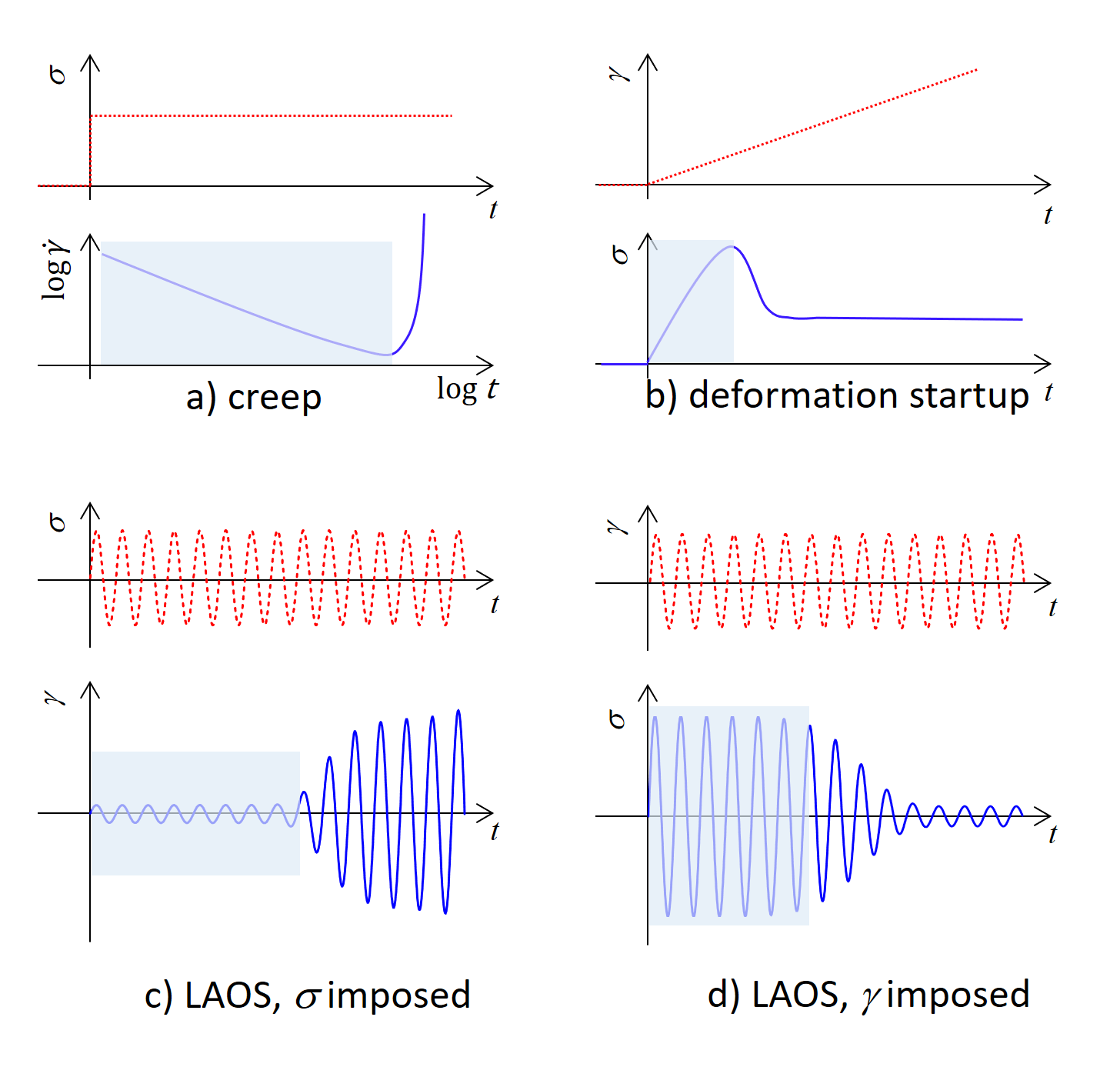}
    \caption{Schematic representation of mechanic failure as probed in various standard rheological tests: $\sigma$, $\gamma$ and $\dot{\gamma}$ are the stress, strain, and strain rate, respectively. LAOS stands for large-amplitude oscillatory strain or stress. For each test, the imposed quantity is shown in the upper panel as a red dashed line, while the measured one is displayed in the corresponding bottom panel as a solid blue line. The shaded boxes show time intervals during which microscopic precursors of failure may be detected.}
    \label{fig:failure_scheme}
\end{figure}

These tests are schematically shown in Fig.~\ref{fig:failure_scheme}. In creep, failure manifests itself by an abrupt increase of the deformation rate, which follows an induction time during which the sample is deformed at a much lower (and often decreasing, as depicted in Fig.~\ref{fig:failure_scheme}a) rate. In deformation startup, failure results in the drop of the associated stress, often after a characteristic peak (``stress overshoot'', Fig.~\ref{fig:failure_scheme}b). In large amplitude oscillatory stress or strain tests (LAOS), failure is signalled by a marked increase of the deformation (Fig.~\ref{fig:failure_scheme}c) or drop of the stress (Fig.~\ref{fig:failure_scheme}d), respectively. Note that in soft matter the stress does not fully vanish after failure, nor does the deformation or deformation rate grow unboundedly, since the system viscosity cannot become smaller than that of the solvent. The shaded boxes in Fig.~\ref{fig:failure_scheme} show the time interval over which microscopic precursors of failure may occur. In order for them to be detected, rheology tests such as those depicted in Fig.~\ref{fig:failure_scheme} should be coupled to a microscopic probe, typically microscopy or scattering. As briefly discussed in Sec.~\ref{sec:techniques}, simultaneous measurements of the microscopic and macroscopic response of soft matter to a mechanical drive have been greatly developed in the past years, both in experiments and computer simulations.

In this Perspective article, we shall focus on soft matter and granular materials, the latter having being studied in pioneering experiments on failure precursors, in particular with light scattering techniques. We emphasize, however, that precursors to material failure are actively investigated also in hard condensed matter. There, acoustic emission measurements have been correlated to mechanical measurements to unveil the physical processes leading to material failure. For instance, Refs.~\cite{deschanel_acoustic_2017,weiss_plastic_2019} discuss the cyclic loading of a polycrystalline piece of metal, during which bursts of acoustic emission reveal the coordinated motions of dislocations in slip bands, which are the precursor of cracks. Similarly, in metallic glasses the acoustic emission activity is a precursor to the stress drop due to shear band propagation~\cite{klaumunzer_probing_2011}. 

The search for precursor signals is also crucial at macroscopic length scales, as in geology.
One of the major challenges remains the prediction of large earthquakes and volcanic eruptions (see e.g. Refs.~\cite{sammonds1992role, hao_accelerating_2017,renard_microscale_2017} and references therein). In this context, we mention insightful laboratory experiments on porous rocks using microtomography, which show evidences of nucleation and propagation of precursory deformation and of self-organization of porosity prior to catastrophic failure~\cite{renard_microscale_2017}.
Other `labquake' experiences determined precursors of failure by measuring the acoustic emission of rock samples under compression~\cite{nataf2014predicting}, finding that in sufficiently porous materials an increase of energy emission occurs near failure. This led to the proposal of using pico-seismicity detection of foreshocks for the prediction of mine collapses in highly porous environments.

On the geological scale, great progress has been made, for example in the field of prediction of glacier failure. Glacier instabilities leading to avalanches are gravity-driven failure phenomena occurring in a natural heterogeneous medium. These events have the potential to cause major disasters, especially when they are at the origin of a chain of processes involving other materials such as snow (snow avalanche), water (flood) and debris (mud flow). The maturation of rupture was shown to be associated with a typical pattern of the time evolution of both the surface velocities and passive seismic activity, signatures that are nowadays in use for warning systems~\cite{faillettaz2019towards}. A last, more exotic yet intriguing example coupling biology and geology is the possible sensitivity of animals to catastrophic events~\cite{bhargava2009earthquake}, like for example the detection by elephants of precursor signals of tsunami~\cite{garstang2009precursor}.

More generally, 'failure' precursors are also of great interest to other domains of research. Apart from the study of fracture in unusual materials, e.g. paper, where an increasing acoustic event rate anticipates the final fracture~\cite{rosti2010statistics}, one can name examples in completely different disciplines: the study of failure in network-models is used, e.g., for the prediction of bank failure~\cite{tam1992managerial}, business and software failure~\cite{ahn2000integrated,tian2005evolutionary}, or even the detection of the onset of heart failure~\cite{choi2016using}. Other domains with interest in the prediction of failure are of course to be found in the context of financial crisis, where for instance collective panic is used to predict crisis~\cite{harmon_anticipating_2015}, and last but not least in the context of health, to understand the ultimate failure --- i.e. death--- of organisms or biological systems, with aging being one type of precursor of failure~\cite{death}.

\section{Simultaneous rheometry and microscopic measurements: experiments and simulations} \label{sec:techniques}

Material failure is, by essence, difficult to predict and subject to large run-to-run variations. Simultaneous measurements of rheological quantities and of the microscopic structure and dynamics during a mechanical test are therefore vital for studying failure precursors. In the last years, various methods have become available to tackle this challenge. Ad-hoc shear cells have been developed for real-space observation~\cite{katsuo_matsuzaka_rheo-optical_1999,wu_new_2007,ober_spatially_2011,chan_simple_2013,boitte_novel_2013,Cohen_RSI2014,tamborini_plasticity_2014,knowlton_microscopic_2014}, typically coupled to a confocal microscope. Real space methods are unsurpassed in that they provide particle-level information on the sample. However, they suffer from limitations in the kind of samples that can be studied, the sample size and their temporal and spatial resolution. Scattering methods are also quite popular, although they don't allow for a direct visualization of the sample. Using coherent radiation and multi-element detectors, it is possible to measure correlators that quantify the structural and dynamical behavior of the material with no need of time average, a key prerequisite for investigating failure. Static and dynamic light scattering and its X-ray equivalent, X-photon correlation spectroscopy (XPCS), have been used coupled to a conventional rheometer~\cite{leheny_rheo-xpcs_2015,ali_rheospeckle:_2016} or with home-made shear cells~\cite{aime_stress-controlled_2016}. Usually, scattering methods provide the structure and the dynamics as a function of scattering vector $q$, i.e. on length scales $\approx 1/q$, and averaged over the whole illuminated sample. In this configuration, it is possible to discriminate the non-affine rearrangements, responsible for plasticity, from the (usually uninteresting) affine deformation that characterizes laminar flow or purely elastic deformation~\cite{leheny_rheo-xpcs_2015,aime_probing_2019a}. Recently introduced variants of the method, by contrast, yield space-resolved data at a single $q$ vector, typically in the form of coarse grained maps of the dynamical activity within the sample~\cite{duri_resolving_2009,pommella_coupling_2019}. The level of coarse graining is of the order of a few hundreds of microns, but the smallest detectable motion is much smaller, of the order of 10 nm or less. Space-resolved light scattering coupled to mechanical tests has also been implemented in the highly multiple scattering limit~\cite{erpelding_diffusive_2008,nagazi_space-resolved_2017,kooij_laser_2018} (diffusing wave spectroscopy, DWS~\cite{weitz_diffusing-wave_1993}). Note that DWS is sensitive to motion on smaller length scales, down to a fraction of nm. Finally, we mention Differential Dynamic Microscopy and other closely related digital Fourier microscopy methods~\cite{giavazzi_digital_2014}, which combine features of both scattering and imaging. While their use coupled to mechanical measurements is still in its infancy~\cite{aime_probing_2019}, they hold a great potentiality, since they can be easily implemented in a standard microscope.

On the theoretical side, numerical modelling of material deformation at the particle level can of course give major insight in the underlying microscopic dynamics leading to failure. Particle-based simulations of amorphous materials, either for dense~\cite{tanguy2006plastic, maloney2006amorphous}, porous~\cite{kun2013approach} or network-forming systems~\cite{bouzid2017network, landrum2016delayed}, allow for a detailed analysis of various observables, from the particle scale to the mesoscopic scale, which can be related to the macroscopic response~\cite{rodney2011modeling}. Despite the advantage of perfect control and easily available information on all scales afforded by particle-based approaches,  allowing notably for the identification of local plastic events \cite{tanguy2006plastic, maloney2006amorphous, kun2014rupture, colombo2013microscopic}, only a few works address the very question of precursors to catastrophic events in terms of a simple observable predictive of the failure time~\cite{cabriolu2019precursors, liu2018creep}. The problem lies in the fact that the finite simulation time and system size impose major limitations to the study of rare events in structurally complex environments. For this reason, mesoscopic models based on an elasto-plastic description of the dynamics under deformation have been developed~\cite{nicolas2018deformation}, aiming at overcoming the limitations intrinsic to particle-level simulations. In these models, the microscopic dynamics are coarse-grained at the mesoscopic level: the  building blocks are of the size of the local rearrangements \cite{rodney2011modeling, nicolas2018deformation}. These models introduce local mesoscopic quantities such as the local stress and deformation, local moduli, and local yield stress or strain criteria, and couple them to the mechanical long range responses to a local rearrangement. They can thus bridge the gap between particle-based simulations and continuum descriptions, while providing information on fluctuations, heterogeneities and other statistical phenomena important for the prediction of precursors towards failure.

\section{Evidence of microscopic precursors} \label{sec:micro_precursors}

\subsection{Granular materials} \label{sec:granular}

\begin{figure}
    \centering
    \includegraphics[width=1\columnwidth]{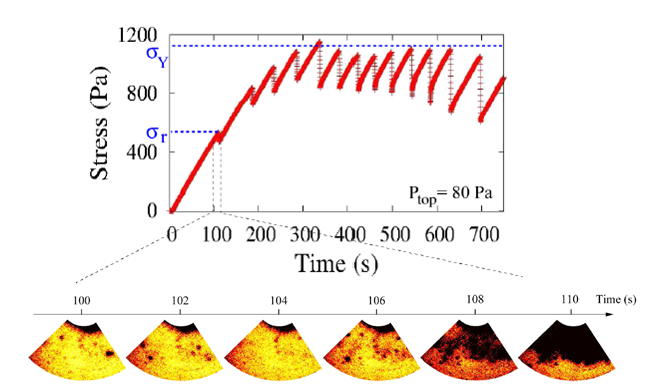}
    \caption{Top panel: stress response of a granular packing submitted to a shear deformation at fixed shear
rate. Bottom panel: map of the microscopic dynamics at the packing surface, obtained with space-resolved diffusing wave
spectroscopy. Dark zones correspond to regions that exhibit plasticity higher than in the yellow ones. Localized ``hot spots'' with high dynamic activity are clearly visible before macroscopic sample yielding. Adapted with permission from~\cite{amon_hot_2012} A. Le Bouil, A. Amon, S. McNamara, and J. Crassous, Physical Review Letters 112, 246001 (2014), http://dx.doi.org/10.1103/PhysRevLett.108.135502. Copyright (2014) by the American Physical Society.}
    \label{fig:granular}
\end{figure}

Pioneering experiments on driven granular materials have been conducted by the group of A. Amon and J. Crassous, who studied a dry packing of glass spheres of size a few hundred of micrometers and typical volume fraction $\sim 60\%$. Thanks to a combination of macroscopic mechanical measurements and light scattering, they have evidenced in several mechanical tests spatially- and temporally-localized zones of intense plastic activity, coined as "hot spots". As an illustration, we show in the top panel of Fig.~\ref{fig:granular} the time evolution of the stress during a shear startup experiment~\cite{amon_hot_2012}, the same kind of experiment as that depicted schematically in Fig.~\ref{fig:failure_scheme}b. The stress exhibits intermittent, sharp drops. By using a space-resolved light scattering method in the multiple scattering regime (DWS), the authors demonstrate the presence of isolated plastic rearrangements~\cite{amon_hot_2012}, which coalesce forming clusters spanning the whole field of view a few seconds before each major stress drop. These localized events, of typical lateral size $15$ beads, can be regarded as precursors of a (transient) macroscopic failure. While in those experiments only the surface of the granular packing is probed, more recent work demonstrates their existence also in the bulk. For example, in Ref.~\cite{amon_experimental_2013} hot spots signing micro-rupture events and corresponding to precursors of avalanches have been identified, thanks to a setup allowing to image the sample from the side.

We note that avalanches are very similar to a stick-slip processes: during stick-slip, localized microscopic rearrangements have been identified by optical imaging and shown to precede macroscopic slip events. Furthermore, the accumulation of these local displacements has been associated with macroscopic creep~\cite{nasuno_friction_1997}. We mention also that avalanche precursors have been evidenced using non-linear acoustic probing in a packing of millimeter-sized glass beads~\cite{zaitsev_pre-avalanche_2008}. Finally, precursors of shear banding, a widespread mode of failure, have been identified when a granular packing is subjected to a biaxial compression~\cite{le_bouil_biaxial_2014, le_bouil_emergence_2014,houdoux_plastic_2018}. Here, space-resolved DWS reveals the emergency of transient, short microbands with a well-defined orientation, different from that of the macroscopic band that will eventually appear. Remarkably, these precursors appear at a deformation as low as half of the deformation where the permanent, macroscopic shear band occurs.

Transient microbands associated with a loss of flow homogeneity, as observed in the experiments by~\citet{le_bouil_emergence_2014}, have also been evidenced in simulations of frictional particles \cite{kuhn1999structured, welker2011precursors, gimbert2013crossover}. These bands consist of localized abrupt rearrangements that induce flow in the surrounding material via non-local, elastic-type interactions~\cite{mcnamara2016eshelby}. Furthermore, a recent numerical particle-based study has shown how frictional dynamics in granular materials favors the nucleation of micro shear cracks that coalesce into system-spanning fracture events when approaching failure~\cite{karimi2019plastic}. So far, however, no simple single observable allowing for the prediction of macroscopic failure could be identified.

An exception is the toy-model recently studied by~\citet{amon2017avalanche}. The model consists of a one-dimensional chain of elastically coupled frictional sliders on an inclined plane, for which a threshold in the number of active sliders has been identified, beyond which an avalanche, equivalent to a failure event, occurs. This threshold delimits the transition from an initial smooth increase of the number of active sliders to an accelerating regime where a dramatic increase of the activity takes place. This can thus be interpreted as a reliable precursor to failure and may serve as a warning signal prior to catastrophic events. However, linking the number of sliders in the toy-model to experimentally measurable observables in more realistic systems remains an open challenge.

\subsection{Soft dense disordered systems}  \label{sec:dense}

What is the nature of the dynamics that lead soft dense materials to failure under an applied external load? This question has stimulated several experimental and theoretical efforts in the last years \cite{rodney2011modeling, bonn2017yield, nicolas2018deformation}. Indeed, there is a large body of literature on the rheology of soft dense disordered systems, such as packings of emulsion droplets, polymer microgels, and colloidal particles. Many of these works aim at a deep understanding of the mechanically-driven solid-to-fluid transition by coupling rheology to additional experimental probes, typically at a mesoscopic or microscopic level. As a first example, we mention the work of Divoux \textit{et al.} on a dense assembly of polymer microgels~\cite{divoux_stress-induced_2011}, whose creep response displays the same sudden upturn of the deformation rate as that depicted schematically in Fig.~\ref{fig:failure_scheme}a. By coupling the rheology test to local velocity profile measurements using ultrasound imaging~\cite{gallot_ultrafast_2013}, the authors evidenced the formation of a series of shear bands that travelled across the sample, before the transition from a solid-like to a fluid-like regime. Note however that these bands were observed in the last regime of the creep experiment, when $\dot{\gamma}$ increases. By contrast, no changes could be detected before or around the minimum of $\dot{\gamma}$, either due to the lack of sensitivity of the ultrasound technique or to the fact that this technique does not give access to the relevant parameter.

Arguably, the non-monotonic behavior of the deformation rate, which initially decreases but eventually grows catastrophically, must be accompanied by some change at the microscopic level, although such precursor may be elusive in dense, amorphous systems, for which no major structural changes may be expected, due to crowding. For some specific systems, however, a structural signature may be accessible. This is the case, e.g., of the dense assembly of anisotropic soft crystallites formed by the surfactant-based hexagonal phase of Ref.~\cite{bauer_collective_2006}, for which a creep experiment was performed simultaneously to structural measurements by small-angle X-ray scattering. The authors of this work measured the evolution of the crystallite orientation, focusing on the initial creep regime, where $\dot{\gamma}$ decreases as a power law with time, before reaching a minimum and growing rapidly, as in Fig.~\ref{fig:failure_scheme}a. The amount of crystallites oriented perpendicular to the shear direction ---thus resisting flow--- was found to continuously decrease and to reach a minimum exactly at the minimum of $\dot{\gamma}$. Consequently, the solid-to-fluid transition could be interpreted as the result of the de-percolation of a network of crystallites oriented perpendicular to the deformation direction and resisting shear.

\begin{figure}
    \centering
    \includegraphics[width=1\columnwidth]{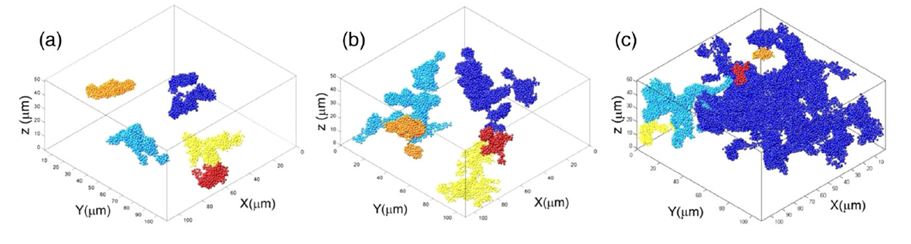}
    \caption{Evolution of clusters of highly mobile particles, as obtained from the analysis of confocal microscopy images taken during a shear startup experiment on a colloidal glass. The strain is $2.1$\% (a), $4.9$\% (b), and $10.1$\% (c). The clusters are found to percolate at yielding (c). Adapted with permission from~\cite{ghosh_direct_2017} A. Ghosh \textit{et al.}, Physical Review Letters 118, 148001 (2017), http://dx.doi.org/10.1103/PhysRevLett.118.148001. Copyright (2017) by the American Physical Society.}
    \label{fig:Clusters}
\end{figure}

The two examples discussed above provide clues about the origin of the solid-to-fluid transition, but unfortunately do not evidence clearly precursors to this transition. In this perspective, coupling rheology to the direct visualization of colloidal particles by confocal microscopy appears as a promising strategy. Moreover, because confocal microscopy allows tracking each individual particle, a comparison with numerical simulations is more straightforward than with other experimental techniques. We illustrate this approach with two recent works, which establish a link between the rheological response and single-particle motion. References~\cite{sentjabrskaja_creep_2015,ghosh_direct_2017} deal with colloidal glasses comprising a dispersion of sterically stabilized particles of micrometer size, at a typical volume fraction of $\sim 60$\%. Crystallization is avoided by using either polydisperse particles~\cite{sentjabrskaja_creep_2015} or a bimodal distribution of particle sizes~\cite{ghosh_direct_2017}. In these experiments, the researchers tracked all the colloidal particles in the field of view and identified those that were most mobile. During shear startup (see the scheme of Fig.~\ref{fig:failure_scheme}b), Gosh~\textit{et al.}~\cite{ghosh_direct_2017} found clusters of particles with enhanced mobility, which grew with time and percolated at yielding, as shown in Fig.~\ref{fig:Clusters}. Similarly, during a creep test, Sentjabrskaj~\textit{et al.} identified localized dynamical heterogeneities due to very mobile particles, and the correlation length of dynamical heterogeneities was shown to increase at the onset of flow. Note that these experiments are in the same spirit as the pioneering measurements by Schall~\textit{et al.}, who unveiled dynamic heterogeneity during shear, although no simultaneous rheology measurements could be performed at that time~\cite{schall_structural_2007}.

Failure and yielding in dense systems have also been extensively studied by computer simulations. Quite generally, soft dense systems exhibit unusual stress-strain curves and complex rheological behaviour, leading to several kinds of interesting out-of-equilibrium transitions, which are accompanied by intermittent dynamics \cite{bonn2017yield,nicolas2018deformation},  and can be regarded as precursors to failure. Here, we will briefly discuss some results focusing on dense assemblies of soft particles (e.g. dense colloidal suspensions, foams or emulsions), which typically yield via a relatively smooth solid-to-fluid transition. However, it is worth mentioning recent advances in the understanding of the opposite situation, where a glass suddenly fails by fracturing. This is the case of the so-called ``ultra-stable'' glasses, created in numerical simulations using smart \textit{ad-hoc} algorithms~\cite{grigera2001fast, berthier2017configurational}. The resulting glass has properties similar to those (extrapolated) for samples aged during an exceptionally long time span. In this respect, they are closer to equilibrium than conventional glasses, which motivates their ``ultra-stable'' designation. By tuning numerically the stability, these systems undergo a transition from soft, ductile materials (for conventional glasses) to brittle glasses that will break suddenly without exhibiting any precursor~\cite{Ozawa6656, PhysRevE.98.040901}, in the ultra-stable limit.

\begin{figure}
    \centering
    \includegraphics[width=1\columnwidth]{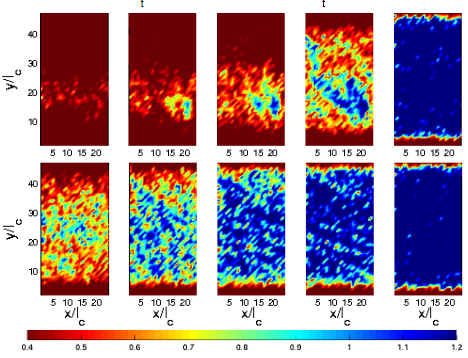}
    \caption{Snapshots of displacement maps obtained in a computer simulation of the shear startup of a glass. The maps show the non-affine component of the local displacement, cumulated from $t=0$ to (from left to right) $t = 7470, 14940, 22410, 29880, $ and $85905$. The top and bottom rows refer to a strongly aged and a young sample, respectively. Reproduced with permission from~\cite{Chaudhuri2013} P. Chaudhuri and J. Horbach, Physical Review E 88, 040301(R) (2013), http://dx.doi.org/10.1103/PhysRevE.88.040301. Copyright (2013) by the American Physical Society.}
    \label{fig:simulation-creep-dyn-het}
\end{figure}

Most of the numerical studies on conventional soft glasses focus on the different kinds of yielding transitions under a constant applied deformation rate, as in Fig.~\ref{fig:failure_scheme}b. Quite generally, dense systems evolve from an elastic regime at small deformations, to a plastic flow regime, beyond the static yield stress~\cite{varnik2004study}. The nature of this transition in the initial, transient regime, potentially associated to avalanche-type dynamics~\cite{combe2000strain, karmakar2010statistical}, has been recently investigated in the context of non-equilibrium phase transitions~\cite{jaiswal2016mechanical,leishangthem2017yielding, ozawa2018random, popovic2018elastoplastic}. The plastic avalanches that occur prior to yielding are often termed precursors to failure. Note however that most studies focus on the statistics of these avalanches, which exhibit generic power-law features as expected for phase transition scenarios. The scale-free nature of theses events clearly does not help in defining a precursor in the form of a simple observable hinting at failure prior to its occurrence.

Theoretical and numerical investigations have also been devoted to delayed failure under an imposed stress, although results from particle-based simulations are still scarce. One of the first investigations in this direction was conducted by Chaudhuri and Horbach, who studied the creep dynamics of a model colloidal system, to which a constant shear stress was imposed through the walls of the simulation box~\cite{Chaudhuri2013}. They demonstrated the existence of spatially heterogeneous dynamics at the onset of flow, as evidenced by the nucleation and growth of mobile regions that may be regarded as precursors of failure~\cite{Chaudhuri2013, shrivastav2016yielding}, see Fig.~\ref{fig:simulation-creep-dyn-het}. It is worth emphasizing the similarity of these results to those obtained in the creep experiments on colloidal glasses by Sentjabrskaja \textit{et al.} mentioned above~\cite{sentjabrskaja2015creep}.

\begin{figure}
    \centering
    \includegraphics[width=1\columnwidth]{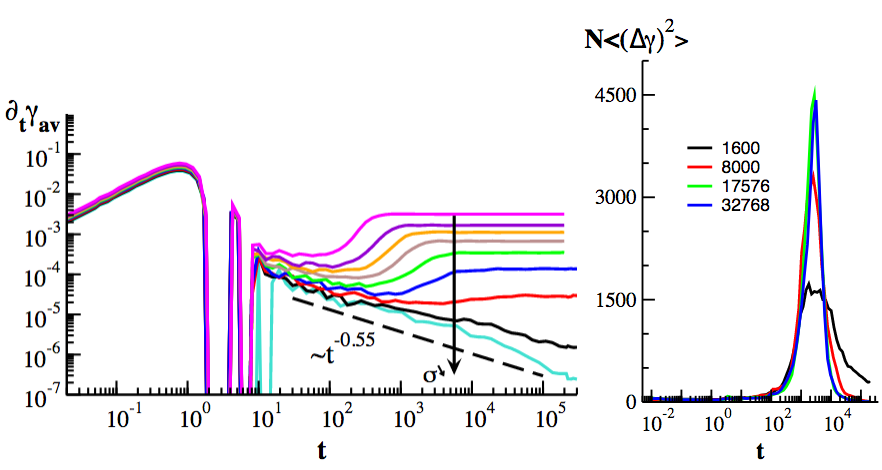}
    \caption{Left panel: Strain rate evolution as a function of time in particle-based simulations of creep of a soft glass. From bottom to top, the stress is (in simulation units) $\sigma = 0.080,
0.085, 0.090, . . ., 0.125$. Right panel: sample to sample fluctuations of the strain as a function of time for different system sizes $N$, as indicated by the label, and for a fixed applied stress $\sigma = 0.1$. The trivial scaling expected for random fluctuations has been factored out by multiplying the fluctuations amplitude by $N$. Adapted from Ref.~\cite{cabriolu2019precursors}.}
    \label{fig:simulation-creep}
\end{figure}

In simulations, wall effects should be dealt with carefully, given the limited sample size. Cabriolu and coworkers have developed a new method~\cite{cabriolu2019precursors} of simulating creep in systems with periodic boundary conditions, which avoids any wall effect and provides the bulk creep dynamics. This method has been used to investigate strain rate fluctuations in the creep dynamics of a colloidal glass model, focusing on the effect of the system size~\cite{cabriolu2019precursors}. As shown in the left panel of Fig.~\ref{fig:simulation-creep}, at sufficiently large applied stress the strain rate exhibits a delayed upturn in the accessible time window, similarly to the situation schematized in Fig.~\ref{fig:failure_scheme}a. The right panel of Fig.~\ref{fig:simulation-creep} displays run-to-run fluctuations of the strain, for $\sigma = 0.1$, the intermediate stress in the left panel (green curve). Remarkably, large fluctuations of $\gamma$ are observed prior to the upturn of $\dot{\gamma}$. The magnitude of these precursors of failure have a non-trivial scaling with system size, as seen in the right panel of Fig.~\ref{fig:simulation-creep}.

As mentioned at the end of Sec.~\ref{sec:techniques}, one of the main problems in particle-based simulations is the limited size of the simulated system. Accordingly, lattice-based models have been introduced, where the relevant mechanisms at play on the microscopic scale are coarse grained on a larger, mesoscopic scale. The key ingredients of these models are typically the occurrence of local plastic rearrangements and their elastic propagation throughout the whole sample. One of the first attempts to study athermal creep dynamics using such an elasto-plastic description is reported by~\citet{bouttes2013creep}, who demonstrated the strong dependence of the creep dynamics on the initial conditions  and the development of shear bands as a possible precursor to yielding. The dynamics associated with creep in a system at finite temperature were explored with similar techniques by~\citet{merabia2016creep}, but without focusing on the possible occurrence of precursors.

\begin{figure}
    \centering
    \includegraphics[width=1\columnwidth]{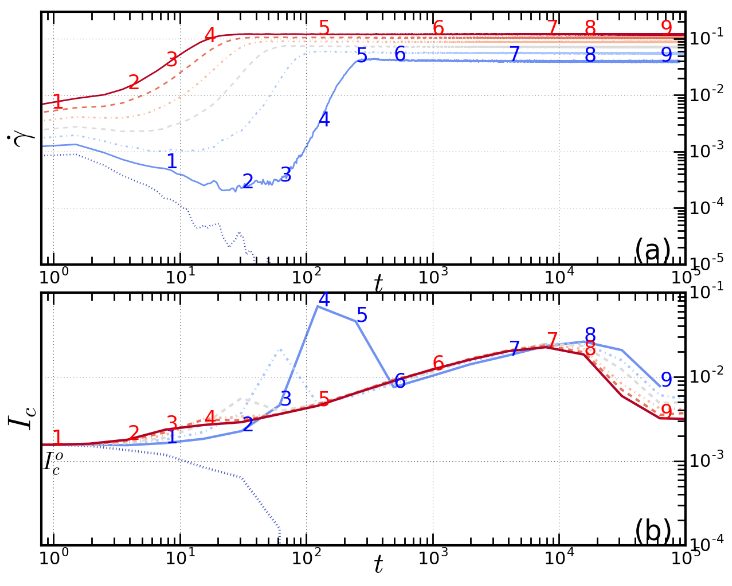}
    \caption{a): Strain rate evolution during creep, as calculated for an elasto-plastic lattice model. b): Corresponding evolution of $I_c$, the intensity of spatial correlations of plastic activity. $I_c$ is defined as the spatially integrated absolute value of the correlation map of plastic events. The red lines correspond to a young sample that exhibits a smooth, monotonic transition towards flow and a modest bump of $I_c$, between points 2 and 4 of the curve. The blue line corresponds to a well-aged sample that shows a characteristic non-monotonic creep curve, accompanied by a peak of $I_c$, corresponding to shear localisation (not shown here). The sudden raise of the correlation intensity can be interpreted as a precursor of the material failure. Adapted from \citet{liu2018creep}.}
    \label{fig:meso-model}
\end{figure}

\citet{liu2018creep}, on the contrary, introduced and studied several quantities exhibiting a non-monotonic behaviour, which may thus be regarded as precursors of yielding. One example is shown in Fig.~\ref{fig:meso-model}b, where $I_c$, a measure of the cooperativity of plastic events, shows a distinct peak in between the points labelled by 2 and 6 of the creep simulation. Remarkably, this precursor occurs concomitantly to the change of regime of $\dot{\gamma}$. Indeed, as seen in Fig.~\ref{fig:meso-model}a, the latter goes through a minimum around phases 2-3 and grows steeply signalling failure between points 3 and 5. Note that delayed failure is clearly seen for aged samples (solid blue curve in Fig.~\ref{fig:meso-model}), while a very smooth increase of the deformation rate is measured for the younger sample (red solid curve), for which only a small bump of $I_c$ is seen, between points 2 and 4. Samples with intermediate age (dashed and dotted lines) lay in between these two extreme cases. Older samples are arguably closer to equilibrium and thus more `stable' than young ones. Therefore, the behavior reported here is strongly reminiscent of the ductile-to-brittle transition mentioned above when discussing ultra-stable glasses.

\subsection{Network-forming systems}  \label{sec:networks}
Network-forming systems, such as colloidal gels formed by the aggregation of attractive particles, or polymer and biological gels and elastomers often display an intriguing behavior when submitted to a constant load. Their macroscopic response exhibits little changes with time, until the material suddenly and catastrophically fails, as schematized in Fig.~\ref{fig:failure_scheme}a. Examples include failure of colloidal, protein, or polymer gels under a constant shear stress~\cite{skrzeszewska_fracture_2010,sprakel_stress_2011,lindstrom_structures_2012,leocmach_creep_2014,landrum2016delayed,aime_microscopic_2018}, onset of fracture in elastomers under tensile strain~\cite{ducrot_toughening_2014,kooij_laser_2018} and hydrogels under a flexion stress~\cite{bonn_delayed_1998}, and even the delayed contraction of actin networks under the internal stress generated by molecular motors~\cite{alvarado_uncovering_2018}. Delayed failure also occurs in weak gels that eventually collapse under their own weight (see e.g.~\cite{buscall_rationalising_2009} and references therein): this delayed sedimentation is of great practical, in addition to academic, interest, because it affects the shelf life time of many products, e.g. in the food, cosmetic and paint industry. Finally, we note that delayed failure may also occur under oscillatory loading, as depicted in Figs.~\ref{fig:failure_scheme}c and d. In these fatigue tests, the sample is submitted to a large number of  deformation cycles at constant amplitude, until it catastrophically fails~\cite{rogers_echoes_2014,perge_time_2014,vandoorn_strand_2018}.

\begin{figure}
    \centering
    \includegraphics[width=1\columnwidth]{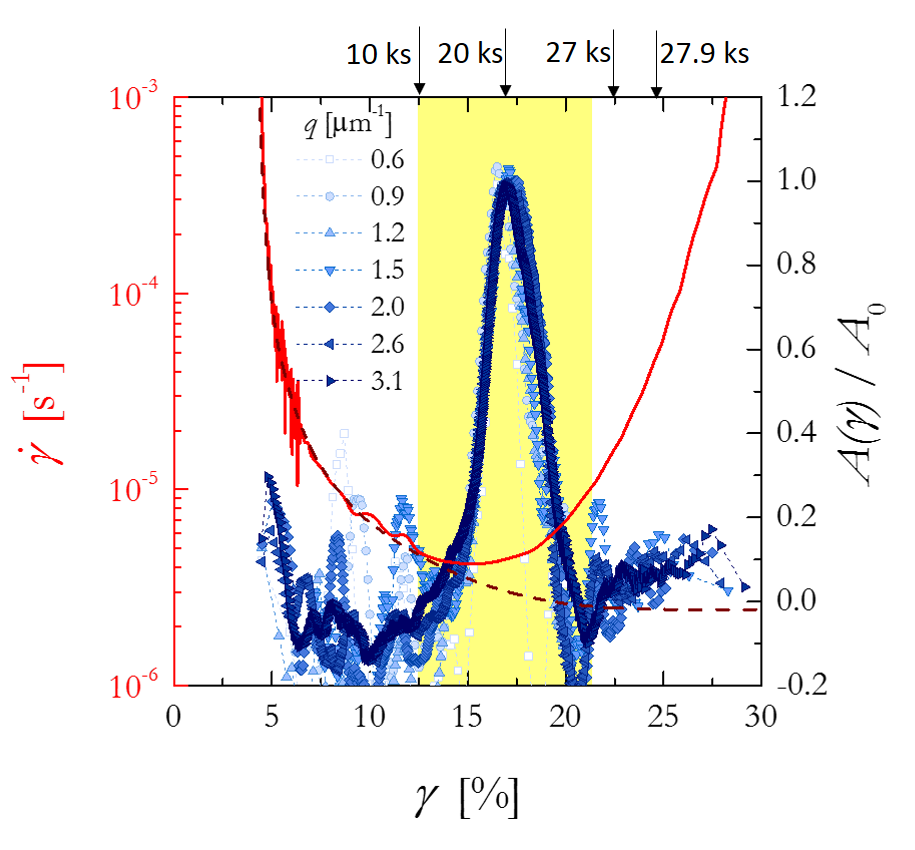}
    \caption{A burst of microscopic dynamics precedes the macroscopic failure of a gel in a creep experiment. Red line and left axis: strain rate as a function of accumulated strain. Dashed line: extrapolation of $\dot{\gamma}$ as obtained from a fit in the linear regime. Blue points and right axis: plastic activity per unit strain increment as obtained form the microscopic dynamics measured at several $q$ values, as shown in the label. The arrows on the top axis show the time $t$ since the start of creep, for various accumulated strains. Adapted from Ref.~\cite{aime_microscopic_2018}.}
    \label{fig:aime_PNAS}
\end{figure}

In some cases, measurable changes in the rheological response precede failure. In creep experiments on various systems~\cite{leocmach_creep_2014,aime_microscopic_2018}, the strain rate initially decreases as a power law with time $t$ since applying a step stress, but eventually goes through a minimum before sharply increasing at failure (see Fig.~\ref{fig:aime_PNAS} for an example). For some systems, such as the protein gel of Ref.~\cite{leocmach_creep_2014}, this feature appears to be remarkably robust: the minimum always occurs roughly at half the failure time, even when the latter is varied over several orders of magnitude by changing the applied stress. In fatigue tests on the same protein gels, the rupture stress $\sigma_c$ can be empirically estimated by monitoring how non-harmonicity builds up in the strain signal for growing $\sigma$, in the low stress ($\sigma < \sigma_c$) regime. While such macroscopic signatures are intriguing, they cannot unveil the microscopic origin of failure. Furthermore, changes in the rheological behavior often occur quite smoothly, limiting their usefulness as a practical means to predict rupture: consider as an example how broad is the minimum of $\dot{\gamma}$ in Fig.~\ref{fig:aime_PNAS} or in Fig. 2b of Ref.~\cite{leocmach_creep_2014}.

As discussed above in reference to dense systems, changes in the microstructure must occur during the latency time preceding failure. Indeed, microstrucutral evolution is frequently invoked as the underlying process that is ultimately responsible for failure, see e.g. Refs.~\cite{sprakel_stress_2011,lindstrom_structures_2012,BartlettSuddencollapsecolloidal2012,leocmach_creep_2014}. However, structural changes are typically quite subtle and thus difficult to detect. In fatigue tests on a colloidal gel coupled to X-ray scattering~\cite{rogers_echoes_2014}, Rogers \textit{et al.} could not detect any change in the structure factor up to failure, although they did observe an enhancement of the microscopic dynamics. In the creep measurements of Ref.~\cite{aime_microscopic_2018}, a slight anisotropy of the scattered light was measured, but it was shown to be simply related to the macroscopic deformation $\gamma(t)$ of the gel network, with no additional information content. Numerical work has shed light on the microscopic plastic events that eventually lead to failure, in both fatigue~\cite{vandoorn_strand_2018} and creep~\cite{landrum2016delayed} tests, highlighting the limited impact that they have on the average structure. For example, Landrum and coworkers have shown that in colloidal gels failure under creep may occur by breaking a surprisingly small fraction of the particle bonds, of the order of 0.1\%~\cite{landrum2016delayed}. This probably explains why experiments in general fail to detect structural precursors of failure, with a few notable exceptions. Bartlett and coworkers were able to follow by confocal microscopy the coarsening of gel strands and the competition between the formation and rupture of particle bonds in a gel that eventually collapsed due to gravity. On mesoscopic length scales, Poon \textit{et al.} unveiled the formation of fractures, using dark field imaging. These fractures evolved into macroscopic channels, eventually leading to gravity-induced large-scale recirculation and the rapid collapse of a colloidal gel~\cite{poon_delayed_1999}. Mesoscopic fractures were also observed concomitantly to the minimum of the strain rate in the experiments on protein gels mentioned above~\cite{leocmach_creep_2014}.

An alternative strategy consist in measuring quantities associated to the system dynamics, rather than its structure. This approach has proved to be quite rewarding. Van der Kooij and coworkers used space-resolved DWS to study the microscopic dynamics around the notch of a pre-damaged elastomer under tensile strain~\cite{kooij_laser_2018}. They observe a localized speed up of the dynamics a few seconds before the onset of crack propagation. It is perhaps not totally unexpected that the dynamics speed up close to failure. Remarkably, however, dynamic precursors may also occur thousands of seconds before failure, as demonstrated by Fig.~\ref{fig:aime_PNAS}. In this experiment~\cite{aime_microscopic_2018}, a step stress is imposed at $t=0$ to a colloidal gel, whose dynamics are measured concomitantly to the macroscopic deformation rate using dynamic light scattering. A burst of irreversibly rearrangements is measured at all probed scattering vectors, which correspond to length scales comparable to the gel mesh size. Using a simple model, the scattering data are converted into a quantity $A(\gamma)$ proportional to the amount of plastic rearrangements per unit volume and unit strain increment. This ``plastic activity'' is shown as blue points in Fig.~\ref{fig:aime_PNAS}: it starts at $t \approx 10^4$ s ($\gamma \approx 12.5\%$) and lasts a few thousands of seconds. These rearrangement events are responsible for the weakening of the network, eventually leading to its failure. Interestingly, the onset of plasticity approximately coincides with the departure of $\dot{\gamma}$ from its initial power-law decay, thus providing a microscopic explanation for the change in the deformation rate. From a practical point of view, it was argued in Ref.~\cite{aime_microscopic_2018} that $A(t)$ is a better predictor of failure as compared to $\dot{\gamma}$, since it changes more abruptly and is thus easier to detect.

Dynamic precursors have also been reported for the protein gels of Ref.~\cite{leocmach_creep_2014}. In this case, the mesoscopic velocity was measured using ultrasonic imaging, again concomitantly to the macroscopic shear deformation in a creep test. The minimum of the deformation rate was found to coincide with the onset of spatio-temporal fluctuations of the local velocity. It is possible that these fluctuations reflect microscopic rearrangements similar to those seen in Ref.~\cite{aime_microscopic_2018}. They may also be the consequence of the propagation of strain fields generated by the nucleation and propagation of cracks, indeed seen outside the region imaged by ultrasounds. Interestingly, similar fluctuations of the velocity maps have also been seen in fatigue tests on a carbon black gel~\cite{perge_time_2014}. While more experiments ---possibly coupled to realistic computer simulations--- will be needed to fully understand the origin of the signal detected in the examples discussed here, it is clear that dynamic precursors are emerging as a powerful tool to detect minute changes in network-forming systems that eventually lead to their failure.

On the theoretical side, one of the simplest approaches to failure in network-forming systems is to consider two material blocks connected by a large number of aligned fibers. These fibers share a global, externally applied load and break irreversibly when their elongation exceeds a randomly distributed threshold; this is the basis of fiber-bundle models. In the thermally activated breakdown of such a model precursors have been identified in the statistics of the energy released prior to global failure~\cite{roux2000thermally}, which has been shown to be related to self-organised criticality \cite{pradhan2001precursors}. In their simplest versions, these models are treated at a mean-field level or for 1D systems. Unfortunately, in this case the models cannot describe the heterogeneous and anisotropic propagation of cracks. Extending this approach to higher dimensions, fuse networks connect lattice nodes by a given conductance that breaks beyond a threshold, thereby burning the fuse. The statistics of the critical dynamics before failure of fuse models has been studied, e.g., by \citet{zapperi2005crack} and for polymer resettable fuses by \citet{cheng2010failure}. These models can capture features of experiments: for example, the acoustic emission due to microfractures in composite inhomogeneous materials has been reported to exhibit a strong analogy with percolation in fuse networks~\cite{garcimartin1997statistical, guarino1998experimental}.

From a mechanical perspective, one can replace the voltages in a fuse model with the local displacement at a given node of masses connected through springs. Using for the energy calculation the Hamiltonian of a network of random springs with a given local stiffness leads thus to the formulation of spring models. Such models have indeed been pioneered by~\citet{de1976relation} to tackle the question of gel formation through cross-linking. By applying a load, precursors can be studied in a similar fashion as for the fuse models, and have found to fall in the same universality class \cite{nukala2005statistical}.

Although the literature on network-forming fuse and spring models is extensive, the definition of failure precursors in the sense of a quantity allowing for the prediction of the failure time has only poorly been addressed. One possible reason could be the deep link between these models and self-organised criticality, which involves scale-free phenomena and thus could rule out the possibility of defining such precursors in the micro-fracture signals prior to yielding.

\section{Outlook and open questions}  \label{sec:outlook}

In the previous sections we have discussed several examples of experiments and simulations that unveiled various kinds of precursors to material failure. In spite of the growing activity in this field, we believe that there are still many key questions waiting for an answer. In this concluding section, we list some of the open issues, chosen among those that seem to us the most interesting and urgent.

In this Perspective article, we have focused on \textit{when} driven soft matter may fail. A related, long-standing question is that of \textit{where} the material will fail. While in crystalline solids plastic rearrangements are directly related to structural defects, identifying a ``defect'' in the structure of an amorphous solid is much more challenging. There is an extensive literature on possible structural defects in disordered environments that could precede the formation of local plastic events. Since this question is not central to this article, we simply mention here a recent work (and the references therein), which compares some of the proposed quantities, such as locally-defined density, potential energy, short range order, lowest shear modulus, participation fraction and yield stresses~\cite{patinet2016connecting}. Finding the most informative parameter is often challenging; an interesting approach recently proposed to circumvent this difficulty consists in using machine-learning methods to establish correlations between the local structure and the occurrence of plastic events~\cite{cubuk_structureproperty_2017}, without choosing \textit{a priori} a specific quantity.

Most of the quantities discussed in Refs.~\cite{patinet2016connecting,cubuk_structureproperty_2017}
require a particle-level knowledge of the the structure, which is not accessible experimentally in many systems of interest. This difficulty is also at the core of the search for precursors of failure discussed in this Perspective article. Microscopy-based experiments can in principle tell us the full story. However, the field of view is limited, while often precursors are localized yet impact the sample over large distances. Moreover, changes in the structure and rearrangement events occurring well ahead failure are typically quite subtle. Thus, a great precision is required in determining the sample structure and its evolution. This can only be achieved for a limited number of specific systems, e.g. micron-sized fluorescent colloids imaged by confocal microscopy. Scattering methods, on the contrary, are quite sensitive to small-scale motion (down to less than 1 nm for DWS and XCPS) and can probe large sample volumes. Moreover, simple scalar quantities can be calculated from dynamic scattering measurements, which are powerful indicators of failure precursors, see e.g. the plastic activity $A$ shown in Fig.~\ref{fig:aime_PNAS}. However, scattering experiments don't provide direct information on the microscopic origin of the events responsible for a rearrangement, e.g. the breakage of a bond in a network-forming system. For gels and networks, real progress is to be expected by coupling scattering measurements to molecular-level probes such as mecanophores, which yield an optical signal related to the amount of stress on a bond or its failure. Mecanophores have been recently included in elastomers~\cite{ducrot_toughening_2014,gostl_pextended_2016,clough_covalent_2016} and DNA hydrogels~\cite{merindol_modular_2019}.

Computer simulations can shed light on the detailed microscopic mechanisms associated to the rearrangement events detected in experiments. However, particle-level simulations suffer from limitations similar to those of microscopy experiments when addressing the spatial-temporal organisation and the coalescense of plastic events towards macroscopic failure. Indeed, the system size is typically relatively small and the range of interparticle interactions that can be implemented for a reasonably large number of particles is restricted. This is especially true for creep dynamics under an externally imposed stress leading to delayed failure, where the slow, out-of-equilibrium dynamics governed by rare plastic events represent a true challenge for simulation algorithms. The limitations given by size and time restrictions are particularly severe for network-forming systems, because the simulation box must span a macroscopic number of mesh sizes, each containing a large number of particles.

The possibility of performing large scale simulations, either using parallel MPI computing on national clusters or using GPGPU computing to run massively parallel implementations on graphic cards allows nowadays to investigate numerically the evolution of systems (with simple interaction potentials) with more than 1 million particles. This is sufficient to investigate phenomena that go beyond the local scale and are not constrained by the system size. For example, the localisation and growth of shear-bands in aged dense systems has been shown to exhibit a cross-over to a non-trivial scaling behavior at large enough system size \cite{vasisht2017emergence}, and large-scale simulations have demonstrated the possibility of the formation of multiple shear bands under an externally imposed shear deformation~\cite{vasisht2018permanent}. For network-forming systems, the available computational power is sufficient to simulate three dimensional systems on the scale of several mesh sizes. This allows for the study of individual plastic events and the resulting effects on the environment, which represents a first important step in the direction of building a multi-scale understanding of failure in complex network environments.

Many of the works discussed here have investigated, both experimentally and numerically context, `simple' systems, such as model colloidal fractal gels, simple elastomers, or dense assemblies of soft particles interacting through a centro-symmetric repulsive potential. Future work on failure precursors will certainly address more complex, real-world systems, including those with exceptional mechanical properties and resilience to failure, in order to better understand the microscopic origin of these properties. Examples range from multiple networks~\cite{varrato_arrested_2012,ducrot_toughening_2014,dimichele_aggregation_2014}, to reinforced elastomers, systems developed for the food industry, and biological or biomimetic materials~\cite{casares_hydraulic_2015,neviaser_basic_2012}.

Other open questions are more fundamental. For example, although several experiments hint at changes in the structure or dynamics of dense systems during the latency period preceding failure, it seems to be difficult to pinpoint simple (e.g. scalar) parameters that unveil failure precursors. Precursors, by contrast, are more easily seen in network-forming systems. At this stage, it is not clear if this difference reflects a fundamental distinction between these two kinds of amorphous soft solids, or it merely stems from the fact that particle rearrangements in densely packed systems are necessarily more restricted than in networks, where more free space is available to their constituents.

In the Introduction, we have schematized several kinds of mechanic tests where failure precursors may be detectable. One open question concerns the relationship (if any) between microscopic failure precursors under different loading conditions. The ultimate goal would be to rationalize delayed failure for different kinds of rheological tests, in the same spirit as that of~\citet{keshavarz_nonlinear_2017}, who recently proposed a failure criterion for network-forming protein gels that applies to both shear startup and stress relaxation tests.

Finally, one fascinating question is that of the generality of failure precursors. Work on several different soft systems suggests intriguing analogies, but it is probably too early to answer this question, let alone to reach a coherent theoretical view that may explain such similarities. At a more general level, we have briefly mentioned failure precursors in hard condensed matter and, beyond material science, in fields as diverse as ecology, medicine and economics. Rationalizing analogies and differences between these research areas appears as a challenging yet potentially highly rewarding endeavor.

\section{Acknowledgements}

We acknowledge financial support from the French CNES and ANR (grant No. ANR-14-CE32-0005, FAPRES), and the EU (Marie Sklodowska-Curie ITN
Supolen Grant 607937). K.~M.~acknowledges financial support from CEFIPRA Grant No. 5604-1 (AMORPHOUS-MULTISCALE).
We are indebted to many colleagues whose contributions, either direct or through enlightening discussions, have been essential to our work in this field. Although it is impossible to list all of them here, we wish to thank in particular S. Mora, C. Ligoure, E. Bertin, S. Aime, J.-L. Barrat, D. Vlassopoulos, T. Divoux, and E. Del Gado.




\balance

\providecommand*{\mcitethebibliography}{\thebibliography}
\csname @ifundefined\endcsname{endmcitethebibliography}
{\let\endmcitethebibliography\endthebibliography}{}

\end{document}